\documentclass[pre,twocolumn,twoside,showpacs,
superscriptaddress,floatfix]{revtex4}
\usepackage{graphics,amssymb,amsmath,epsfig}

\begin{document}
\title{Finite-size scaling of synchronized oscillation on complex networks}

\author{Hyunsuk Hong}
\affiliation{Department of Physics and RINPAC, Chonbuk National University,
Jeonju 561-756, Korea}

\author{Hyunggyu Park}
\affiliation{School of Physics, Korea Institute for Advanced Study, Seoul
130-722, Korea}

\author{Lei-Han Tang}
\affiliation{Department of Physics, Hong Kong Baptist University, Kowloon Tong,
Hong Kong SAR, China}

\date{\today}

\begin{abstract}
The onset of synchronization in a system of random frequency
oscillators coupled through a random network is investigated.
Using a mean-field approximation, we characterize sample-to-sample
fluctuations for networks of finite size, and derive the
corresponding scaling properties in the critical region. For
scale-free networks with the degree distribution $P(k)\sim
k^{-\gamma}$ at large $k$, we found that the finite size exponent
$\bar{\nu}$ takes on the value $5/2$ when $\gamma>5$, the same as
in the globally coupled Kuramoto model. For highly heterogeneous
networks ($3<\gamma <5$), $\bar{\nu}$ and the order parameter
exponent $\beta$ depend on $\gamma$. The analytic expressions for
these exponents obtained from the mean field theory are shown to
be in excellent agreement with data from extensive numerical
simulations.
\end{abstract}
\maketitle

\section{Introduction}

The popularity of complex networks in the description of
interactions among individuals in various biological and social
contexts has inspired theoretical studies of ordering phenomena on
networks in recent years~\cite{ref:review_Dorogovtsev}.
The small-world properties
of such networks, as emphasized first by Watts and
Strogatz~\cite{ref:WSnetworks}, suggest that a simple mean-field (MF)
description of the ordering transition is often
appropriate~\cite{ref:WSnetworks,ref:review_Dorogovtsev}.
More complicated situations may arise as in, e.g., scale-free
networks with a low degree exponent, where heterogeneity in
the network topology smears out the transition
significantly~\cite{ref:hetero-DGM,ref:FSS-HHP,ref:Potts_Igloi,ref:Ising_Leone,ref:Percol_Havlin,ref:DP_SFN}.
In general, randomness in network connections, which can be considered
as a form of quenched disorder, can have a profound effect on the
ordering process and the ensuing scaling behavior. This is a topic
in the network research which has not been sufficiently explored so far.

Synchronization of coupled oscillators is a representative dynamical problem
on complex networks~\cite{ref:review_Dorogovtsev}. By varying the coupling
strength $J$ among the oscillators, various dynamical phenomena
can be observed, ranging from independent oscillators at $J=0$ to
fully synchronized state at $J>J_f$. The desynchronization
threshold $J_f$ is an important property in many applications and
its dependence on the network topology has been investigated in
great detail~\cite{ref:synch_networks}.  For coupled oscillators
with a distribution of intrinsic
frequencies~\cite{ref:Winfree,ref:Kuramoto,ref:Pikovsky,ref:Kiss},
a finite fraction of the population can remain entrained in
frequency even in the presence of a large number of run-away
oscillators. The entrained cluster of oscillators disappears only
at a (much) lower coupling strength $J_c$. Near the entrainment
threshold $J_c$, strong fluctuations in various static and dynamic
properties of the system are expected, as in usual critical
phenomena.

The entrainment transition on scale-free
networks with the degree distribution $P(k)\sim k^{-\gamma}$ has
been treated analytically by several
groups~\cite{ref:SFN_mf,ref:synch_SFN,ref:MarylandGroup}. It has
been shown that, in the infinite size limit, the transition is
expected at a finite coupling strength for $\gamma>3$, while the
entrained cluster persists at any nonzero coupling strength for
$\gamma\leq 3$. The critical exponent $\beta$ describing vanishing
behavior of the order parameter on the supercritical side is shown
to be equal to $1/2$ for $\gamma>5$ and $1/(\gamma-3)$ for
$3<\gamma<5$~\cite{ref:synch_SFN}.

For systems of finite size, the entrainment transition becomes
blurred and rounded over a range of the coupling strength.
In addition, randomness in the network
topology, as well as the random choice of oscillator frequencies,
introduces sample-to-sample variations in the entrainment threshold.
A full description of the finite size effects, which requires
a detailed characterization of dynamic fluctuations in specific samples,
is currently not available~\cite{ref:Strogatz,ref:Acebron}.
Fortunately, in the case of the globally coupled Kuramoto model
where the mean-field theory works well on the supercritical side,
the sample-to-sample fluctuations of the order parameter can be
characterized analytically~\cite{ref:Hong_Entrainment}.
Comparison with numerical simulations indicates that temporal
fluctuations of the order parameter only play a subdominant role
in the broadening of the transition region due to finite
size~\cite{ref:Hong_Entrainment,ref:HPT-BC}.
The success of this approach suggests a novel procedure
to derive finite-size scaling (FSS) relations under a MF approximation.

In the present paper, we extend the above MF treatment to
coupled random frequency oscillators on complex networks.
Unlike the globally coupled case, the MF equations derived here
are not expected to be exact due to the finite connectivity
of individual vertices on the network. Nevertheless, as we show below,
the FSS exponents obtained depend only on certain general properties
of the network. We also present results from extensive
simulations on uncorrelated scale-free networks. The numerically determined
values of the exponents as a function of the degree exponent $\gamma$
of the network agree well with the analytic predictions.
Our study indicates that the FSS at the entrainment transition on
uncorrelated scale-free networks is also governed by
fluctuations in the distribution of intrinsic oscillator frequencies,
with temporal order parameter fluctuations playing a less dominant role.

The paper is organized as follows. In Sec. II we introduce the
dynamic model and derive the mean-field equations. Section III
contains a treatment of the mean-field equations in the
neighborhood of the entrainment transition. A finite-size scaling
form for the order parameter is obtained. Results from numerical
simulations of the model are presented in Sec. IV and compared
with the analytic predictions. Section V contains a brief summary
of our results.

\section{The Model and Mean-field Equations}

\subsection{The model}

An undirected network of $N$ vertices is defined by the
adjacency matrix $\{a_{ij}\}$, where
$a_{ij}=1$ if two vertices $i$ and $j$ are connected and 0
otherwise. The degree of a vertex $i$ is the number of vertices
connected to $i$, denoted by $k_i=\sum_j a_{ij}$.
To each vertex $i$ we associate an oscillator whose dynamics
is described by the equation of motion for the phase $\phi_i$,
\begin{equation}
\dot\phi_i = \omega_i - J\sum_{j=1}^N a_{ij} \sin(\phi_i -
\phi_j), \label{eq:KM_onSFN}
\end{equation}
where $\omega_i$ is the intrinsic frequency of the oscillator.
The second term on the right-hand side of
Eq.~(\ref{eq:KM_onSFN}) denotes coupling to neighboring
oscillators on the network with a positive strength ($J>0$).

In this paper, the oscillator frequencies $\omega_i$ in a given sample
are drawn independently from a distribution $g(\omega)$ which is
assumed to be a smooth function and symmetric about its maximum
at $\omega=0$.
In addition, we shall limit ourselves to random networks with no
degree-degree correlation among neighboring vertices.
An algorithm that generates such a network with no self
linking nor multiple links between vertices is discussed in
Ref. \cite{ref:Catanzaro}. The numerical results presented in Sec. IV
are for scale-free networks generated using the static
model of Ref. \cite{ref:SNUstatic_SFN}.

\subsection{The mean-field approximation}

At sufficiently strong coupling, the system described by
Eq.~(\ref{eq:KM_onSFN}) exhibits a synchronization phenomenon
where a finite fraction of oscillators in the system become
entrained in frequency. Transition to the random state at a
critical coupling strength $J_c$ has been considered at the
mean-field level by several
authors~\cite{ref:SFN_mf,ref:synch_SFN,ref:MarylandGroup}. To
understand the general idea behind such an approach, let us first
introduce a set of instantaneous local fields defined by
\begin{equation}
H_i e^{i\theta_i} \equiv \sum_j a_{ij} e^{i\phi_j},
\label{eq:localfield}
\end{equation}
where $H_i\geq 0$ denotes the amplitude and $\theta_i$ the phase,
respectively. Eq.~(\ref{eq:KM_onSFN}) can now be written in a more
suggestive form,
\begin{equation}
\dot\phi_i = \omega_i - H_i J\sin(\phi_i - \theta_i).
\label{eq:localfield_eq}
\end{equation}

As usual, the mean-field approximation decouples the set of $N$
dynamical equations by replacing $H_i$ and $\theta_i$ with
suitable time-averaged quantities which are then determined in a
self-consistent manner. Unlike the globally coupled Kuramoto
model, such a mean-field treatment is not exact due to the finite
connectivity of individual oscillators on the network, so dynamic
fluctuations are not averaged out even for an infinite system.
However, provided the network is sufficiently well connected
(as compared to fragmentation into local ``communities''), and
because of its small-world properties, once a cluster of entrained
oscillators is formed, it will affect the dynamics of all
oscillators in the system through a {\em global} ``ordering
field'' $H\exp(i\theta)$. The precise effect of the global field
on individual oscillators is subject to renormalization by the
local community of a given oscillator. For a well-connected
network, it is reasonable to expect that renormalization of
oscillator's response function does not change qualitatively
results of the mean-field theory where such effects are ignored.
This is to be confirmed by numerical investigations.

With the above caveat, let us proceed to the derivation of the
mean-field equations, with particular emphasis on the finite size
effect. The key approximation we introduce is to replace each phase
factor $e^{i\phi_j}$ in the sum of Eq.~(\ref{eq:localfield}) by the
global ordering field $H\exp(i\theta)$ acting through the edge
connecting vertices $i$ and $j$. Equation (\ref{eq:localfield_eq})
now becomes
\begin{equation}
\dot\phi_i = \omega_i - k_iHJ \sin(\phi_i - \theta).
\label{eq:mf_eq}
\end{equation}

Consider a steady-state situation with a constant $H$
and a linearly advancing $\theta=\Omega t$, where $\Omega$
is the phase velocity of the entrained cluster.
From Eq.~(\ref{eq:mf_eq}), the oscillator at vertex $i$ is entrained with
\begin{equation}
\phi_i=\theta+\sin^{-1}[(\omega_i-\Omega)/(k_iHJ)]
\label{eq:phi-entrained}
\end{equation}
if $|\omega_i-\Omega|\le k_iHJ$,
and detrained otherwise. In the latter case,
the time averaged value of $e^{i(\phi_i-\theta)}$ is given by
\begin{equation}
\overline{e^{i(\phi_i-\theta)}} = i\Bigl(\frac{\omega_i-
\Omega}{k_iHJ}\Bigr) \Bigl[1
-\sqrt{1-\Bigl(\frac{k_iHJ}{\omega_i-\Omega}\Bigr)^2}\Bigr].
\label{eq:avg_phase}
\end{equation}
Here and elsewhere the overline bar denotes time average.

The self-consistent equations for $H$ and $\Omega$ are obtained by
setting $H$ equal to the average of
$\overline{e^{i(\phi_i-\theta)}}$ over all edges of the network.
Since each vertex $i$ contributes $k_i$ times to the average, we
may write
\begin{equation}
H=\frac{\sum_{i} k_i\overline{e^{i(\phi_i-\theta)}}}
{\sum_{i}k_i}= \frac{1}{N}\sum_{i=1}^N \frac{k_i}{\langle
k\rangle} \overline{e^{i(\phi_i-\theta)}}.
\label{eq:H-def}
\end{equation}
Here $\langle k\rangle$ is the average degree of a vertex in the
network. Substituting Eqs.~(\ref{eq:phi-entrained}) and
(\ref{eq:avg_phase}) into Eq.~(\ref{eq:H-def}), and separating out
real and imaginary parts, we obtain
\begin{eqnarray}
H&=&\frac{1}{N}\sum_{i=1}^N \frac{k_i}{\langle k\rangle}
\sqrt{1-f_i^2}
\Theta(1-|f_i|),\label{eq:H-eq}\\
0&=& \sum_{i=1}^N(\omega_i-\Omega)
\Bigl[1-\sqrt{1-f_i^{-2}}\Theta(|f_i|-1)\Bigr],
\label{eq:Omega}
\end{eqnarray}
where $f_i=(\omega_i-\Omega)/(k_i HJ)$, and $\Theta(x)$ is the
Heaviside step function which takes the value 1 for $x\geq 0$ and
$0$ otherwise.

Equations (\ref{eq:H-eq}) and (\ref{eq:Omega}) are our mean-field
equations for a given network of $N$ oscillators with a particular
set of intrinsic frequencies $\{\omega_i\}$. They are
invariant under a uniform shift of all oscillator frequencies
(with a corresponding change in $\Omega$).

At any value of $J$, Eqs.~(\ref{eq:H-eq}) and (\ref{eq:Omega})
admit a trivial solution $H=0$ (with arbitrary $\Omega$). In the
next section, we shall analyze nontrivial solutions that appear
above the entrainment threshold $J_c$, with particular attention
on sample-to-sample variations when $N$ is finite.

\section{Solution of the mean-field equations}

From Eqs.~(\ref{eq:H-eq}) and (\ref{eq:Omega}) it is obvious that
the solution for $H$ depends not only on the coupling strength $J$
but also on the particular choice of the intrinsic frequencies
$\{\omega_i\}$ as well as the particular network topology in a
given sample. For sufficiently large $N$, it is possible to give a
statistical description of the sample-to-sample fluctuations. As
we show below, the analysis also yields the finite-size scaling of
the critical properties in the neighborhood of the entrainment
transition. By symmetry we shall seek a solution at $\Omega=0$ and
ignore weak finite-size corrections which do not alter our main
conclusions.

\subsection{Self-averaging in the infinite size limit}

To proceed, let us write Eq.~(\ref{eq:H-eq}) in a symbolic form,
\begin{equation}
H=\tilde\Psi(H),
\label{eq:Psi-self-consistent}
\end{equation}
where
\begin{equation}
\tilde\Psi(H)\equiv \frac{1}{N}\sum_{i=1}^N \frac{k_i}{\langle
k\rangle} \sqrt{1-\Bigl(\frac{\omega_i}{k_i HJ}\Bigr)^2}
\Theta\Bigl(1-\frac{|\omega_i|}{k_iHJ}\Bigr). \label{eq:Psi-tilde}
\end{equation}
Terms in the sum can be grouped according to their degree $k_i$.
When the network size $N\rightarrow\infty$,
the number of terms in each group at a given $k$ grows linearly
with $N$, and hence the self-averaging over the distribution
$g(\omega)$ is expected. This consideration leads to the result,
\begin{equation}
\Psi(H)\equiv\lim_{N\rightarrow\infty}\tilde\Psi(H)
=\langle\tilde\Psi(H)\rangle =\frac{1}{\langle k\rangle}\sum_{k}
P(k) k u(kHJ). \label{eq:Psi-H}
\end{equation}
Here $P(k)$ is the degree distribution of vertices on the network, and
\begin{equation}
u(x)=\int_{-x}^xd\omega g(\omega)\sqrt{1-\omega^2/x^2}
\label{eq:u-x}
\end{equation}
is a monotonically increasing function of $x$ which approaches 1
as $x\rightarrow\infty$. For small $x$, $u(x)\simeq
\frac{\pi}{2}g(0)x +\frac{\pi}{16} g''(0)x^3$.

We now consider the behavior of $\Psi(H)$ at small $H$, assuming
$\langle k^2\rangle=\sum_k k^2P(k)$ to be finite
(i.e., $P(k)$ falls off faster than $k^{-3}$ at large $k$).
To facilitate the analysis, we write
\begin{equation}
u(x)=ax-\hat{u}(x),
\label{eq:hat-u}
\end{equation}
where $a=\frac{\pi}{2}g(0)$ and $\hat{u}(x)\simeq -{\pi\over
16}g''(0)x^3$ for small $x$. Substituting Eq.~(\ref{eq:hat-u})
into Eq.~(\ref{eq:Psi-H}), we obtain
\begin{equation}
\Psi(H)=a\frac{\langle k^2\rangle}{\langle
k\rangle}JH-\hat\Psi(H), \label{eq:Psi-H-new}
\end{equation}
where
\begin{equation}
\hat{\Psi}(H)=\frac{1}{\langle k\rangle}\sum_{k} P(k) k
\hat{u}(kHJ). \label{eq:hat-Psi}
\end{equation}
The functional form of $\hat{\Psi}(H)$ at small $H$ depends on the
tail of the degree distribution $P(k)$. If $P(k)$ falls faster
than $k^{-5}$ at large $k$, we may use the small $x$ expansion of
$\hat{u}(x)$ to obtain
\begin{equation}
\hat{\Psi}(H)\simeq c_0\frac{\langle k^4\rangle}{\langle
k\rangle}(JH)^3, \label{eq:hat-Psi-gamma-large}
\end{equation}
where $c_0=-({\pi/16})g''(0)$ is a positive constant.
On the other hand, if $P(k)\simeq Ak^{-\gamma}$ at large $k$
with $3<\gamma<5$, $\langle k^4
\rangle$ diverges and the above expansion becomes invalid.
Instead, contributions to the sum in Eq.~(\ref{eq:hat-Psi}) come
mainly from vertices with $k\sim (JH)^{-1}$. Since the fraction of
vertices in this degree range is proportional to $k^{-\gamma+1}$,
and each contributing an amount $k$ to the sum, we estimate
$\hat{\Psi}(H)\sim (JH)^{\gamma -2}$ in this case. More precisely,
replacing the sum over $k$ by an integral, we obtain
\begin{eqnarray}
\hat{\Psi}(H)&\simeq& \frac{1}{\langle k\rangle}\int_0^\infty dk
P(k) k \hat{u}(kHJ)
\nonumber\\
&\simeq&\frac{1}{\langle k\rangle}\int_0^\infty dk
Ak^{-\gamma+1}\hat{u}(kHJ)
\nonumber\\
&=&c_1 (JH)^{\gamma-2}. \label{eq:hat-Psi-gamma-small}
\end{eqnarray}
Here $c_1=A{\langle k\rangle}^{-1}\int_0^\infty dx
x^{-\gamma+1}\hat{u}(x)$ is another positive constant.

\subsection{Sample-to-sample variations in a finite network}

To work out the statistics of the sample-to-sample fluctuation
$\delta\tilde\Psi(H)\equiv\tilde\Psi(H)-\Psi(H)$, we note that
$\tilde\Psi(H)$ can be viewed as the mean value of the random
variable,
\begin{equation}
\eta(\omega, k)= \frac{k}{\langle k\rangle}
\sqrt{1-\Bigl(\frac{\omega}{kHJ}\Bigr)^2}
\Theta\Bigl(1-\frac{|\omega|}{kHJ}\Bigr)
\end{equation}
over $N$ realizations.
From the central limit theorem, we expect $\delta\tilde\Psi(H)$ to
satisfy the Gaussian distribution with zero mean and a variance
given by
\begin{equation}
\langle(\delta\tilde\Psi)^2\rangle=\frac{1}{N}\bigl(\langle\eta^2\rangle
-\langle\eta\rangle^2\bigr)\equiv \frac{D(H)}{N}.
\label{eq:deltaPsi-var}
\end{equation}
As $\langle\eta\rangle=\Psi(H)\sim H$ at small $H$, we only need
to focus on
\begin{equation}
\langle\eta^2\rangle=\sum_k P(k)\frac{k^2}{\langle k\rangle^2}
\int_{-kHJ}^{kHJ}d\omega
g(\omega)\Bigl[1-\Bigl(\frac{\omega}{kHJ}\Bigr)^2\Bigr]
\end{equation}
close to the transition. If $P(k)$ falls faster than $k^{-4}$ at
large $k$, the above expression can be easily evaluated at small
$H$ to give
\begin{equation}
D(H)\simeq \langle\eta^2\rangle\simeq \frac{4\langle
k^3\rangle}{3\langle k\rangle^2}g(0)JH. \label{eq:D-gamma-large}
\end{equation}
On the other hand, if $P(k)\simeq Ak^{-\gamma}$ with $3<\gamma<4$,
we may replace the sum over $k$ by an integral which yields
\begin{equation}
D(H)\simeq d_1 (JH)^{\gamma-3}, \label{eq:D-gamma-small}
\end{equation}
where $d_1=A\langle k\rangle^{-2}\int_0^\infty du u^{2-\gamma}
\int_{-u}^u d\omega g(\omega)(1-\omega^2/u^2)$ is another positive
constant.

Summarizing the above results, in the neighborhood of the entrainment
transition, we may write the self-consistent equation
(\ref{eq:Psi-self-consistent}) for $H$ in the form,
\begin{equation}
H=(J/J_c)H-c(JH)^p+d(JH)^{q/2}N^{-1/2}\xi,
\label{eq:Psi-mean-fluc}
\end{equation}
where $c$ and $d$ are positive constants, and $J_c=2\langle
k\rangle/(\pi g(0)\langle k^2\rangle)$ (mean-field value) is the
critical coupling strength at the transition. The term $\xi$ is a
Gaussian random variable with zero mean and unit variance that
represents the combined effect of random frequency and network
realizations in a given sample. The exponents $p=3$ and $q=1$ when
the degree distribution $P(k)$ decays sufficiently fast at large
$k$. For power-law distributions $P(k)\sim k^{-\gamma}$, $p$
switches to the value $\gamma-2$ for $\gamma<5$. Similarly, $q$
switches to the value $\gamma-3$ for $\gamma<4$.

\subsection{Finite-size scaling}

The network size $N$ enters Eq.~(\ref{eq:Psi-mean-fluc}) through
the last , where $\xi$ varies from sample to sample. At $J=J_c$, a
positive $\xi$ (50\% of the samples) yields a solution $H\sim
N^{-1/(2p-q)}$, quite a bit larger than the value $N^{-1/2}$ due
to dynamic fluctuations on the detrained side~\cite{ref:Daido}. In
addition, one can easily verify that the term $\xi$ affects the
solution significantly when $J$ is within a distance of order
$N^{-(p-1)/(2p-q)}$ from $J_c$. These properties are summarized in
the scaling solution to Eq.~(\ref{eq:Psi-mean-fluc}),
\begin{equation}
H=N^{-\beta/\bar{\nu}}f\bigl((J-J_c)N^{1/\bar{\nu}}\bigr),
\label{eq:H-N-scaling}
\end{equation}
where the exponents $\beta=1/(p-1)$ and $\bar{\nu}=(2p-q)/(p-1)$.
Unlike the usual FSS expression for pure systems, the scaling function
$f(x)$ varies from sample to sample.

Using the above values for $p$ and $q$, we obtain the following
results for the critical exponents in case of uncorrelated scale-free networks,
\begin{equation}
(\beta,~\bar\nu) = \left\{
\begin{array}{cccl}
\frac{1}{2}, &\frac{5}{2}, & \qquad\mbox{$\gamma>5$;} \\
\frac{1}{\gamma-3}, &\frac{2\gamma-5}{\gamma-3}, & \qquad\mbox{$4<\gamma<5$;}\\
\frac{1}{\gamma-3}, &\frac{\gamma-1}{\gamma-3}, & \qquad\mbox{$3< \gamma < 4$.}
\end{array}
\right. \label{eq:beta_bar}
\end{equation}

In simulation studies, the order parameter
\begin{equation}
\Delta=\frac{1}{N}\sum_{i=1}^N \overline{e^{i(\phi_i-\theta)}}
\label{eq:Delta-def}
\end{equation}
is usually measured. From the solution to the mean-field equation
(\ref{eq:mf_eq}), we obtain
\begin{equation}
\Delta=\frac{1}{N}\sum_{i=1}^N \sqrt{1-\Bigl(\frac{\omega_i}{k_i
HJ}\Bigr)^2} \Theta\Bigl(1-\frac{|\omega_i|}{k_iHJ}\Bigr).
\label{eq:Delta-exp}
\end{equation}
Following the same procedure as above in the calculation of
$\Psi(H)$, we obtain at small $H$,
\begin{equation}
\Delta\simeq a\langle k\rangle JH +d_2 (JH)^{1/2} N^{-1/2}\tilde\xi,
\label{eq:Delta-result}
\end{equation}
where $d_2=[(4/3)g(0)\langle k\rangle]^{1/2}$ and $\tilde\xi$ is another Gaussian random variable. The finite-size term here is negligible even
at the transition. Hence the scaling behavior of $\Delta$ is the same as
that of $H$.

We note that the expressions for the FSS exponent $\bar\nu$ in
Eq.~(\ref{eq:beta_bar}) differ from those obtained by
Lee~\cite{ref:synch_SFN} based on a cluster analysis but without
considering sample-to-sample fluctuations. In fact, $\bar\nu$
obtained here is always larger than that in \cite{ref:synch_SFN}
for any $\gamma$. This observation suggests that the broadening of
the transition region by the random nature of oscillator
frequencies and network connectivity is more significant than
other effects such as those considered in Ref.
\cite{ref:synch_SFN}.

\begin{figure}
\includegraphics[width=0.45\textwidth]{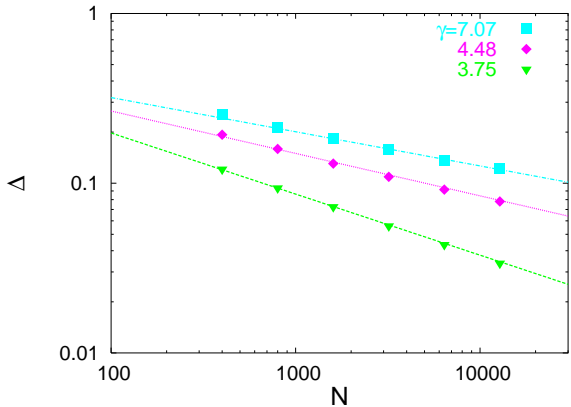}
\caption{\label{fig:Delta} (Color online) Critical decay of
$\Delta$ at the critical point $J_c$ as a function of $N$ for
$\gamma=7.07, 4.48, 3.75$, where $J_c=1.86(2), 1.57(4), 1.17(3)$,
respectively. Lines are drawn with slopes of $-\beta/\bar\nu$
given by the theoretical prediction, Eq.~(\ref{eq:beta_bar}).
 }
\end{figure}

\begin{figure}
\includegraphics[width=0.45\textwidth]{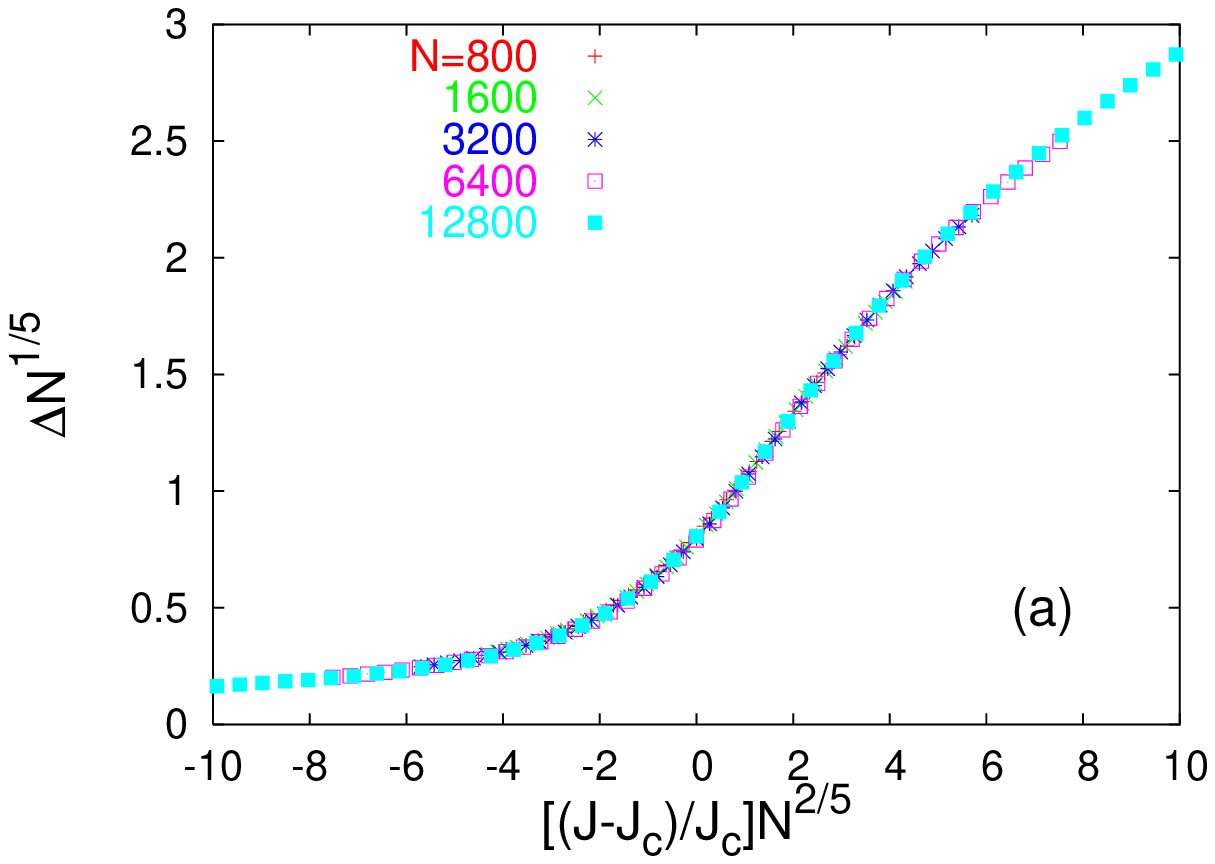}
\includegraphics[width=0.45\textwidth]{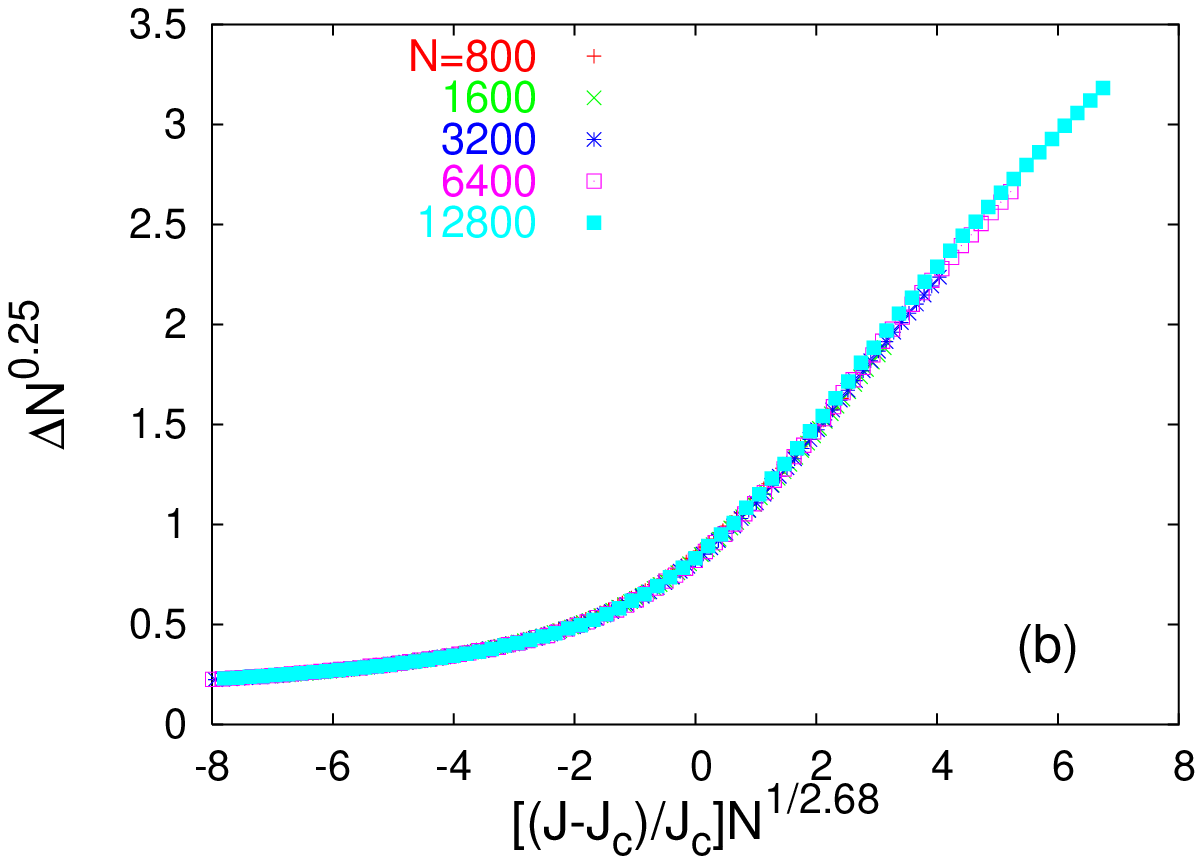}
\includegraphics[width=0.45\textwidth]{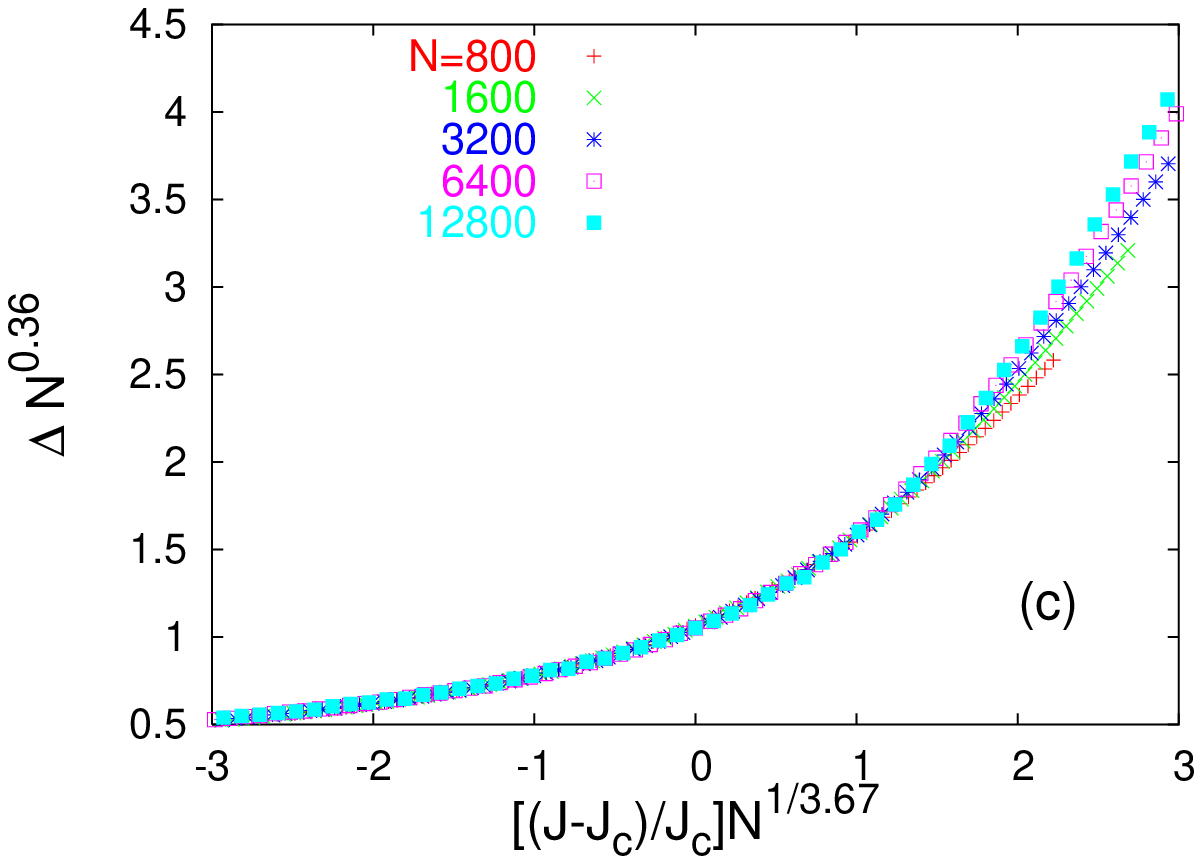}
\caption{\label{fig:collapse} (Color online) Scaling plot of
$\Delta$ for $N=800,1600, \cdots, 12800$ with (a) $\gamma=7.07$,
using $\beta/\bar{\nu}=1/5$ and $\bar{\nu}=5/2$,  (b)
$\gamma=4.48$, using $\beta/\bar{\nu}=0.25$ and $\bar{\nu}=2.68$,
and (c) $\gamma=3.75$, using $\beta/\bar{\nu}=0.36$ and
$\bar{\nu}=3.67$. }
\end{figure}

\begin{figure}
\includegraphics[width=0.45\textwidth]{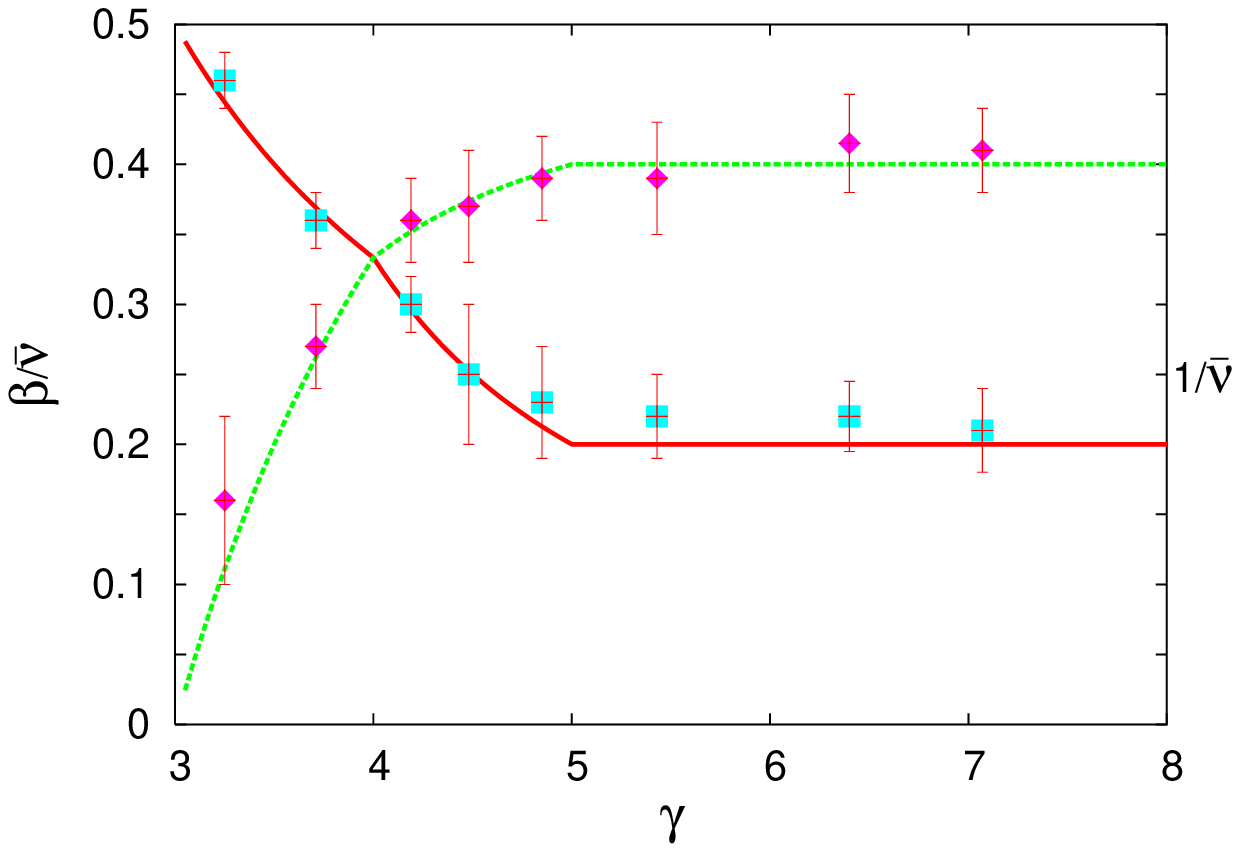}
\caption{\label{fig:betabarnu} (Color online) $\beta/\bar\nu$
(box) and $1/\bar\nu$ (triangle) are plotted as a function of the
degree exponent $\gamma$, showing a good consistency with the
theoretical prediction represented by the full line
($\beta/\bar\nu$) and the dotted one ($1/\bar\nu$), respectively.}
\end{figure}

\section{Numerical results}

To test the validity of our MF analysis, we have performed
extensive numerical simulations on the system governed by
Eq.~(\ref{eq:KM_onSFN}). Scale-free networks used in the
simulation are generated following Ref.~\cite{ref:SNUstatic_SFN},
up to a system size $N=12800$. The intrinsic frequencies of
oscillators are drawn independently from the Gaussian distribution
$g(\omega)=(2\pi\sigma)^{-1/2} \exp(-\omega^2/2\sigma^2)$ with
unit variance ($\sigma^2=1$). The Heun's method~\cite{ref:Heun},
with a discrete time step $\delta t=0.01$, is used to integrate
numerically Eq.~(\ref{eq:KM_onSFN}). Typically, the motion is
followed for $N_t=4\times 10^4$ time steps, with the initial
condition $\phi_i=0$ for all $i$. Data from the first $N_t/2$
steps are discarded in measuring the time average of various
quantities of interest. For each network size, approximately
$10^4$ independent runs are performed, with different realizations
of the intrinsic frequencies as well as network connectivity to
obtain sample averages.

To characterize the entrainment transition numerically, we have
focused on the order parameter defined by
Eq.~(\ref{eq:Delta-def}), or more precisely,
$\Delta=\langle\overline{\Delta_t}\rangle$, where
$\Delta_t=|\frac{1}{N}\sum_{j=1}^N\exp(i\phi_j)|$ at time $t$ and
$\langle\cdots\rangle$ denotes the sample average for a given
size. The critical coupling strength $J_c$ is estimated from the
crossing point in the plot $\Delta N^{\epsilon}$ versus $J$,
varying the exponent $\epsilon$. This value for $J_c$ is checked
against crossing of the Binder cumulants at different size defined
by
$B_{\Delta}=1-\frac{1}{3}\langle\overline{\Delta_t^4}/\overline{\Delta_t^2}^2\rangle$~\cite{ref:HPT-BC},
and compared with the behavior of the dynamic susceptibility
$\chi=N\langle\overline{\Delta_t^2}-\overline{\Delta_t}^2\rangle$.
At the critical point $J=J_c$, the order parameter $\Delta$
exhibits a power law: $\Delta\sim N^{-\beta/\bar\nu}$, which
provides an alternative way for checking the critical value as
well as determining the exponent ratio $\beta/\bar\nu$. With the
values of $\beta/\bar\nu$ so obtained, we estimate the value of
the FSS exponent $\bar\nu$ from the scaling plots of $\Delta
N^{\beta/\bar\nu}$ against $(J-J_c)N^{1/\bar\nu}$ for a broad
range of systems sizes $N$, adjusting $\bar\nu$ to achieve the
best data collapse in the critical region.

Figure~\ref{fig:Delta} shows the critical decay of the order
parameter $\Delta$ at $J_c$ as a function of $N$ on log-log scale.
The critical values of $J$
are given by $J_c=1.86(2), 1.57(4)$, and $1.17(3)$ for
$\gamma=7.07, 4.48$, and $3.75$, respectively.
Each set of data points at a given $\gamma$ fall nicely on a
straight line corresponding to the anticipated power law behavior
$\Delta(J_c)\sim N^{-\beta/\bar\nu}$, and
the slope shows very good agreement with the predicted value
for $\beta/\bar\nu$ according to Eq.~(\ref{eq:beta_bar}).

Figure~\ref{fig:collapse} shows
the scaling plots of the order parameter $\Delta$ for various
system sizes at three different values of $\gamma$. In each case,
the theoretically predicted values given by
Eq.~(\ref{eq:beta_bar}) are used to scale the horizontal and
vertical axes. The data collapse is nearly perfect for
$\gamma=7.07$ and $4.48$, and satisfactory for $\gamma=3.75$.

Finally, we present in Fig.~\ref{fig:betabarnu}
results for $\beta/\bar\nu$ and $\bar\nu$
determined following the procedure described above at various values
of $\gamma$. Error bars are obtained from uncertainties in
the procedure. The two lines correspond to our MF predictions.
The agreement is generally good.

\section{Conclusions and discussion}

In this paper, we investigated the effect of fluctuations
in the frequency distribution on the entrainment transition
of a system of coupled random frequency oscillators.
Self-consistent equations for the global ordering field are derived
for any given oscillator population on a complex network.
Statistical properties of the solution to these equations in
an ensemble of such networks of oscillators are determined.
The analysis enables us to derive a finite-size scaling expression
for the entrainment order parameter. Comparison with numerical
integration of dynamical equations on scale-free networks shows that
the mean-field description correctly captures the finite-size scaling
behavior exhibited by the sample-averaged order parameter $\Delta$.
The values of exponents $\beta$ and
$\bar{\nu}$ which best describe the numerical data show excellent
agreement with the predicted ones from the mean-field theory.

For scale-free networks with a degree exponent $\gamma>5$, our
result for the exponents $\beta=1/2$ and $\bar\nu=5/2$ is the same
as the globally coupled Kuramoto model. Randomness in the network
connection does not appear to affect these values. On this ground
we expect the result to apply to randomly connected networks with
a bounded degree, to the Erd\"os and R\'enyi
network~\cite{ref:Erdos}, as well as to the small-world networks
of Watts and Strogatz~\cite{ref:WSnetworks}. However, the finite
connectivity of oscillators on the network implies that the
dynamic fluctuations are not averaged out and can at least
renormalize the entrainment threshold $J_c$ given by the
mean-field theory. In this regard, analytic derivation of the
sample-dependent, self-consistent equation
(\ref{eq:Psi-mean-fluc}) beyond the mean-field approximation would
be desirable.

This work was supported by Korea Research Foundation Grant funded by
the Korean Government (MOEHRD) (KRF-2006-331-C00123) and by the
Research Grants Council of the HKSAR through grant 202107.

\end{document}